\documentclass[12pt]{article}

\usepackage{mathrsfs}
\usepackage{amsmath}
\usepackage{amssymb}
\usepackage{amsfonts}
\usepackage{latexsym}
\usepackage{graphicx}
\usepackage{latexsym}
\usepackage{color}
\usepackage{indentfirst}

\usepackage[hyperindex]{hyperref}
\hypersetup{colorlinks=true,linkcolor=blue,citecolor=blue,urlcolor=blue}

\def\ln{\,\mbox{ln}\,}

\def\al{\alpha}
\def\be{\beta}
\def\ga{\gamma}
\def\Ga{\Gamma}
\def\de{\delta}
\def\De{\Delta}
\def\ep{\epsilon}
\def\vp{\varepsilon}

\def\th{\theta}

\def\la{\lambda}

\def\si{\sigma}

\def\rh{\rho}

\def\ta{\tau}

\def\ph{\varphi}

\def\na{\nabla}
\def\pa{\partial}

\newcommand{\rd}{\mathrm{d}}

\def\beq{\begin{eqnarray}}
\def\eeq{\end{eqnarray}}

\newcommand{\nn}{\nonumber}

\usepackage{float} 
\usepackage{titlesec}

\titleformat*{\section}{\large\bfseries}
\titleformat*{\subsection}{\normalsize\bfseries}

\begin{document}

\begin{center}

{\Large On the Renormalization in Conformal Quantum Gravity}

 \vskip 6mm

\textbf{Ioseph L. Buchbinder} $^{a,b}$
\footnote{E-mail address: \ buchbinder@theor.jinr.ru},
\ \
\textbf{Petr M. Lavrov} $^{b}$,
\ \
\textbf{Thomas M. Sangy} $^{c,d}$
\footnote{E-mail address: \   thomas.sangy@edu.ufes.br},
\ \
\textbf{Ilya~L.~Shapiro} $^{d,c}$
\footnote{
E-mail address: \   ilyashapiro2003@ufjf.br}
%

\vskip 4mm

a) Bogoliubov Laboratory of Theoretical Physics,
\\
Joint Institute for Nuclear Research,
141980, Dubna, Russia
\vskip 2.5mm

b) Center of Theoretical Physics,
Tomsk State Pedagogical University,
\\
634061, Tomsk, Russia \vskip 1mm
\vskip 2mm

c) PPGCosmo, \ Universidade Federal do Esp\'{i}rito Santo,
\\
29075-910, \ Vit\'{o}ria, \ ES, \ Brazil
\vskip 2mm

d) Departamento de F\'{\i}sica, ICE, Universidade
Federal de Juiz de Fora,
\\
36036-900, Juiz de Fora, Minas Gerais, Brazil
\end{center}
\vskip 4mm

\begin{center}

{\large\bf Abstract}

\begin{quotation}

\noindent
One-loop divergences in classically conformal theory in curved
spacetime is a nontrivial issue if the theory under consideration
possesses gauge invariance. In this case, quantization involves
introducing the gauge-fixing term and the action of ghosts, both
of which are not conformal. The formal proof of the conformal
invariant renormalizability in this situation, including interacting
theories, has been given in the paper from 1984 by one of the
present authors. Owing to BRST symmetry, the contributions of
the gauge-fixing and ghost sectors to the conformal variation of the
effective action cancel each other. This cancellation does not hold
in the finite part of the one-loop and higher-loops effective action,
that results in the  anomaly. In this paper, we extend the early
results on conformal matter fields in external gravitational field
and apply them to conformal quantum gravity, including the case
of conformal gravity coupled to other conformal matter fields.
The gauge-fixing in the Weyl-squared gravity is more complicated,
nevertheless, the proof of the one-loop conformal invariant
renormalization using BRST symmetry is possible. As a
preliminary to conformal quantum gravity, we also present a
detailed general proof of conformal invariant one-loop divergences
in the corresponding semiclassical theory.
\vskip 3mm

\vskip 2mm

\end{quotation}
\end{center}

\newpage


\tableofcontents


\section{Introduction}
\label{Intro}

Local conformal symmetry is one of the cornerstones of semiclassical
gravity. One of the main points is that its violation through the
trace (conformal) anomaly \cite{CapDuf-74,ddi,duff77} provides a
simple and elegant description of quantum contributions to the
vacuum effective action in the high-energy limit. These contributions
underlie such important  applications as Hawking radiation
\cite{black}, the complete formulation of the Starobinsky
inflationary model \cite{fhh,star} (see the recent paper
\cite{StabInstab} for further details and references), and a variety
of other related issues \cite{birdav, BOS,duff94,PoImpo}.
The intuitive arguments (see, e.g., \cite{OUP}) and direct
calculations in all known semiclassical models show that the
one-loop divergences in a classically conformally invariant theory
preserve this invariance.
In particular, it was shown in \cite{Adler:1976jx}, that the
non-renormalized trace of the vacuum energy-momentum tensor
for free gauge field is conformal, regardless the symmetry is
violated by the quantization procedure.

A formal proof of the
conformal invariance of one-loop divergences, valid for both free
and interacting theories conformal coupled to an external
gravitational field, was briefly presented only in the paper
\cite{tmf} within the BRST formulation of quantum gauge
theories, where some issues were not considered in
details\footnote{There is a large literature on the study of
divergences in free conformal scalar, spinor and vector fields
in curved spacetime (see, e.g., \cite{birdav,BOS,OUP} and
references therein). Here we focus on interacting theories in
external gravitational fields.}
The one-loop
divergences can be directly related to the trace anomaly
\cite{duff77}. In this way, conformally invariant one-loop
renormalization leads to the classification of the terms appearing
in the conformal anomaly \cite{ddi,DeserSchwimmer} and to the
derivation of the anomaly-induced effective action in semiclassical
gravity \cite{rie,frts84}.
This form of the effective action is particularly convenient for
applications and is equivalent \cite{Deser1999} to an alternative
form derived independently by solving the heat-kernel equations
\cite{bavi90,Avramidi89,Barvinsky1994}.

It is important to establish a formal proof of conformal invariance
in one-loop renormalization, even though in many cases the result
is already known from direct calculations. There are several reasons
for this. First, it is only natural to seek an understanding of why
a given property holds, rather than merely knowing the result itself.
Another important point is that, in more complicated conformal
theories, there may be uncertainty regarding the correctness of the
existing results. In such cases, a formal proof can provide a reliable
criterion for assessing their validity.
In this respect, the proof presented in \cite{tmf} shows
in general terms that conformal invariance is not expected to persist
beyond the one-loop level. In this way, it provides a deeper
understanding of the conformal anomaly and its possible role in
physical applications.

There is the much less explored case of conformal quantum gravity,
either pure or coupled to conformal matter. The renormalizability
of the general fourth-derivative quantum gravity has been proved in
\cite{Stelle77}. It is worth noting that there are more recent works
\cite{Barvinsky2018,Lavrov:2022ceg} which use more sophisticated
techniques for the proof and cover more general versions of quantum
gravity, with both local and nonlocal actions. On the other hand,
conformal quantum gravity is expected to be
\textit{nonrenormalizable} in higher loops because the trace anomaly
produces violation of conformal symmetry. However, even the
one-loop renormalizability is a non-trivial issue, in both pure
conformal gravity and for the theory including conformal matter
fields.

In the case of pure quantum gravity, i.e., the theory based on the
square of the Weyl tensor, several calculations have been performed,
starting from the seminal classical work \cite{frts82} (see also
\cite{avbar86} and \cite{Ohta-Handbook} for a recent review),
in which the existing
framework for conformal quantum gravity theories in $4D$ was
established. After a number of subsequent verifications
\cite{antmot,Weyl,Ohta2015}, there is now a consensus that
conformal invariance is indeed preserved in the one-loop
divergences. The same conclusion was reached in the original work
\cite{frts82}, but only after employing a special transformation of
the background metric \cite{ETG76,fradvilk78}, sometimes referred
to as ``conformal regularization''. We briefly review this procedure
in the Appendix. In fact, this procedure is not necessary, since
conformal invariance is already present, as demonstrated by
subsequent calculations.

In the conformal quantum gravity coupled to conformal quantum
matter, to the best of our knowledge, only one calculation has been
reported \cite{BuSh86}. Let us note the qualitative difference with
\cite{BKSVW}, where the renormalization structure in grand
unification - type theories (GUTs) coupled to quantum higher
derivative gravity has been considered (see also later work with an
alternative calculation of the same beta functions in
\cite{Agravity}). Its conclusion was that one-loop renormalizability
requires ``conformal regularization'' \cite{ETG76,fradvilk78,frts82}.
This procedure can be readily extended to include scalar fields.
However, the non-conformal result obtained through direct
calculations may be due to a technical mistake, as pointed out in the
recent work \cite{Jack2020}. This example further highlights the
importance of establishing a general proof of conformal invariance
in quantum gravity, including theories coupled to conformal matter
fields.

In the present work, we begin by formulating a technically simpler
proof of one-loop conformal renormalization in semiclassical gravity,
i.e., in a classically conformally invariant theory of matter fields
in an external gravitational background. The simplification arises
from a more direct use of BRST
symmetry~\cite{Becchi-1:1975nq,Becchi-2:1975nq,Tyutin:1975qk},
including the
introduction of the third, i.e., Nakanishi-Lautrup, ghost field.
More importantly, these relatively minor modifications enable us
to extend the proof presented in \cite{tmf} to conformal quantum
gravity, including the case of quantum gravity coupled to quantum
matter.

The paper is organized as follows. In Sec.~\ref{sec2}, we briefly
review the necessary background on local conformal symmetry,
including the actions of the basic conformal fields, the conformal
Noether identities, and the algebra of symmetries involving
diffeomorphism and conformal transformations. In Secs.~\ref{sec3}
and \ref{sec4}, we present, respectively, general arguments
supporting the conformal invariance of the one-loop divergences
and a formal proof based on the BRST formalism. Sec.~\ref{sec5}
contains the
central result of the present work, namely the proof of one-loop
conformal renormalizability in quantum conformal gravity. In
Sec.~\ref{sec6}, this analysis is extended to theories including
matter fields. Sec.~\ref{sec7} discusses the implications for the
conformal anomaly and the anomaly-induced effective action in
quantum gravity. Finally, in Sec.~\ref{Conc}, we present our
conclusions.

\section{Classical theories coupled to gravity
and conformal symmetry}
\label{sec2}

The quantum considerations presented in the following sections
concern the one-loop renormalizability of conformal matter theories
and conformal gravity  within the framework of dimensional
regularization. Therefore, we begin with a brief review of classical
conformal symmetry in theories under consideration in a spacetime
of dimension $n \neq 4$. Quantum aspects will be addressed
whenever appropriate.
Our interest lies in theories of conformal matter fields
$\Phi=\Phi^A$, including generally covariant scalar, spinor, and
vector fields, i.e., $\Phi^{A} = (\ph,\,\psi, \, A_\mu)$, in the
external gravitational field described by the metric $g_{\mu\nu}$.
Whenever a field appears as an argument of a functional, its indices
are omitted; for example, in the present case, we write
$g = g_{\mu\nu}$.

The local conformal transformations read
\beq
g_{\mu\nu}={\bar g}_{\mu\nu}e^{2\si}
,\quad
\Phi^A = {\bar \Phi}^A e^{d_A\si}
,\quad
\si = \si(x).
\label{conftrans}
\eeq
In what follows, we use dimensional regularization, hence
the conformal weights $\,d_A\,$ of the fields
$\,\Phi^A\,$ depend on the dimension $n$. Requiring the
free-field actions to be conformally invariant, one finds
\beq
d_{\text{scalar}} = \frac{2-n}{2}
,\qquad
d_{\text{spinor}} = \frac{1-n}{2}
,\qquad
d_{\text{vector}} = 0.
\label{weights}
\eeq
It is worth noting that the formulation of a conformal vector field
models in an arbitrary dimension $n$ is a nontrivial issue (see, e.g.,
\cite{ETG76,DN84,OsbornStergiou2016,ALW} and references therein).

The infinitesimal form of the transformations (\ref{conftrans})
looks like,
\beq
\de_{c} g_{\mu\nu} = 2g_{\mu\nu}\si
,\quad\,\,
\de_{c} \ph = - \frac{n-2}{2}\,\ph \si
,\quad\,\,
\de_{c} \psi = -   \frac{n-1}{2}\,\psi \si
,\quad\,\,
\de_{c} A_\mu = 0.
\label{inftrans}
\eeq
Here we choose to preserve the transformation law of the vector
field $A_\mu$ (Abelian or non-Abelian) in the same form as in
four-dimensional spacetime, although other, more complicated
possibilities exist \cite{ALW}.

The actions of the free massless fields $S_k$, \ $k = 0,1/2,1$,
are of the form
\beq
&&
S_0\,=\,\int \rd^nx \sqrt{-g}\,\Big\{
\frac12 g^{\mu\nu}\pa_\mu \ph  \pa_\nu\ph
+ \frac12 \xi R \ph^2 \Big\},
\nn
\\
&&
S_{1/2}\,=\,i\int \rd^nx \sqrt{-g}\, {\bar \psi}\ga^\mu \na_\mu \psi,
\nn
\\
&&
S_{1}\,=\,-\frac14 \int \rd^nx \sqrt{-g}\, F_ {\mu\nu} F^{\mu\nu}.
\label{freeactions}
\eeq
In the scalar case, $\xi$ is the nonminimal (coupling) parameter.
The special value
\beq
\xi_{ij} \,=\, \frac{n-2}{4(n-1)}\,\de_{ij} \,,
\label{xin}
\eeq
that provides the conformal invariance of a multiscalar
massless theory in an arbitrary dimension $n$. The conformal
invariance of the vector-field action in arbitrary dimensions can be
achieved in several nontrivial ways \cite{DN84,ALW} by introducing
suitable modifications to the classical action. In what follows,
however, we assume that the vector model is described by
(\ref{freeactions}).

Under the mentioned assumptions, the violations of conformal
invariance in the classical Noether identity are proportional to
$n-4$. Thus, we have
\beq
\mathcal{C}_n S_k
\,=\,
\frac{1}{\sqrt{-g}}\,\Big\{
2g_{\mu\nu}\frac{\de}{\de g_{\mu\nu}}
\,+\, d_k \Phi^A \frac{\de}{\de \Phi^A} \Big\}\,S_k
\,=\,(n-4)\,F_{k} ( \Phi, g ).
\label{Noether}
\eeq
For free fields, $F_{k} ( \Phi, g )$ may vanish for scalars and
spinors, whereas it is nonzero for vector fields unless the special
formulations of \cite{ETG76,DN84,ALW} are employed.

It is easy to verify that the identity (\ref{Noether}) also holds
for the quartic scalar self-interaction, the Yukawa interaction
between scalar and spinor fields, and gauge interactions,
including those involving scalar, spinor, and gauge vector fields.

In the vacuum sector of four-dimensional semiclassical gravity,
renormalizability requires the inclusion of four-derivative
metric-dependent terms in the action (see, e.g., \cite{BOS,OUP}
and references therein). In $n=4$, these terms include a covariant
integral of the Weyl-squared term (the $c$-invariant). For the sake
of generality, we present its $d$-dimensional expression,
\beq
C^2(d)\,=\,C_{\mu\nu\al\be}(n)\,C^{\mu\nu\al\be}(n)
\,=\,R_{\mu\nu\al\be}R^{\mu\nu\al\be} - \frac{4}{d-2}
R_{\mu\nu}R^{\mu\nu} + \frac{2}{(d-1)(d-2)}\,R^2.
\label{W2}
\eeq
Here $d$ may be equal to $4$, to $n$, or take some other value, as
will be discussed in Sec.~\ref{sec7}. Strict conformal invariance of
$\int \rd^nx \sqrt{-g}\,C^2(d)$ holds only in the case $d=n=4$, of
course.

There are also two $N$-terms that are not strictly conformally
invariant but nevertheless satisfy the Noether identity
(\ref{Noether}). The first of these is the covariant integral of
the Gauss-Bonnet term (the Euler characteristic),
\beq
E_4\,=\,R_{\mu\nu\al\be}R^{\mu\nu\al\be}
- 4 R_{\mu\nu}R^{\mu\nu} + R^2,
\label{E4}
\eeq
and the second is the $\Box R$ term. The nonconformal local
term of the same dimension is the integral of $R^2$. If this
term is omitted, the classical vacuum action takes the form
\beq
S_{\text{conf}, \, \text{vac}}\,=\,\int \rd^nx \sqrt{-g}\,
\big\{ a_1C^2 +  a_2E_4 +  a_3\Box R\big\}\,,
\label{actionCQG}
\eeq
and satisfies the classical conformal Noether identity
(\ref{Noether}). The same action (\ref{actionCQG}) provides
a starting point for conformal quantum gravity,
which is the primary subject of the present work.

Finally, in order to quantize conformal gravity, it is necessary
to consider the algebra of the generators associated with its two
symmetries, namely diffeomorphism invariance and local
conformal invariance. The infinitesimal form of these
transformations is
\beq
&&
\de_\xi g_{\mu\nu}
\,=\, \na_\mu \xi_\nu +  \na_\nu \xi_\mu
\,=\, g_{\mu\al}  \pa_\nu \xi^\al + g_{\nu\al}  \pa_\mu \xi^\al
+  \xi^\al \pa_\al g_{\mu\nu}\,,
\nn
\\
&&
\de_c g_{\mu\nu} \,=\, 2  g_{\mu\nu} \si(x)\,,
\label{infis}
\eeq
corresponding to the diffeomorphism
$x^\al = x^{\prime \al} + \xi^\al$ and conformal (\ref{conftrans})
transformations. The conformal transformation does not affect the
coordinates, i.e.,  $\de_c \xi^\al = 0$. Consequently,
we have $\de_c \xi_\mu = \de_c (g_{\mu\al}\xi^\al) = 2\si \xi_\mu$.
Applying these two transformations in different orders, we obtain
\beq
&&
\de_c \de_\xi g_{\mu\nu}
\,=\, \de_c\big( g_{\mu\al}  \pa_\nu \xi^\al
+ g_{\nu\al}  \pa_\mu \xi^\al
+  \xi^\al \pa_\al g_{\mu\nu}\big)
\nn
\\
&&
\qquad \qquad
\,=\, 2\si  g_{\mu\al} \pa_\nu \xi^\al
\,+\, 2\si  g_{\nu\al} \pa_\mu \xi^\al
\,+\,  \xi^\al \pa_\al \big( 2\si g_{\mu\nu}\big)
\label{2trans-1}
\eeq
and
\beq
&&
\de_\xi \de_c  g_{\mu\nu}
\,=\, \de_\xi \big( 2\si  g_{\mu\nu}\big)
\,=\, \de_\xi \tilde{g}_{\mu\nu}
\,=\, \tilde{g}_{\mu\al}  \pa_\nu \xi^\al
+ \tilde{g}_{\nu\al}  \pa_\mu \xi^\al
+  \xi^\al \pa_\al \tilde{g}_{\mu\nu}
\nn
\\
&&
\qquad \qquad
\,=\, 2\si  g_{\mu\al} \pa_\nu \xi^\al
\,+\, 2\si  g_{\nu\al} \pa_\mu \xi^\al
\,+\,  \xi^\al \pa_\al \big( 2\si g_{\mu\nu}\big)\,.
\label{2trans-2}
\eeq
The comparison of the last two formulas shows that
\beq
&&
\big[ \de_\xi ,\, \de_c \big] g_{\mu\nu} \,=\, 0\,.
\label{2trans-com}
\eeq
This means that the generators of the two symmetries form a
closed algebra. Consequently, the corresponding theory can be
quantized using the standard Faddeev-Popov procedure.

\section{Semiclassical theory I. General considerations}
\label{sec3}

The main object of study in this paper is the quantum effective
action
$\Ga[\Phi,g]=\Ga_{\text{fin}}[\Phi,g]+ \bar{\Ga}_{\text{div}}[\Phi,g]$,
where $\Ga_{\text{fin}}[\Phi,g]$ is a finite part after removing
regularization and $\bar{\Ga}_{\text{div}}[\Phi,g]$ is a divergent part.
In the framework of loop expansion and dimensional regularization,
the divergent part has the following structure
\beq
\bar{\Ga}_{\text{div}}[\Phi,g]
\,=\,
\sum_{L=1}^{\infty}\hbar^{L}\bar{\Ga}_{\text{div}}^{(L)}[\Phi,g]
\label{divact}
\eeq
and
\beq
\bar{\Ga}_{\text{div}}^{(L)}[\Phi,g]
\,= \,\frac{1}{\vp^{L}}\,\tilde{\Ga}^{(L)}[\Phi,g]\,.
\label{L-div}
\eeq
Here, $\vp=(4\pi)^2(n-4)$ is a parameter of dimensional regularization
and $\tilde{\Ga}^{(L)}[\Phi,g]$ is a finite functional defining the
$L$-loop contribution to the divergent part of effective action.

At the quantum level, local conformal symmetry is known to be
broken by the trace anomaly \cite{CapDuf-74} (see \cite{duff94}
for a review and numerous references). Many important results in
semiclassical gravity, e.g., Hawking radiation and cosmological
applications \cite{fhh}, including the Starobinsky model \cite{star}
(particularly in its complete form), are related to the anomaly.

The anomaly is directly related to the UV divergences and, for this
reason, it can be relatively simple to study its structure.
Regardless anomaly violates conformal symmetry already in the
finite part of the one-loop effective action, the one-loop,
dimensionally regularized,
functional $\tilde{\Ga}_{\text{div}}^{(1)}$ in classically
conformal theories satisfies the identity (\ref{Noether}),
\beq
\mathcal{C}_n \tilde{\Ga}^{(1)}_{\text{div}}
\,=\, E_{n}^{(1)},
\label{Noether-div}
\eeq
where $E_{n}^{(1)}$ is a functional of the one-loop order
and $\mathcal{C}_n$ is a generator of conformal transformations
defined in (\ref{Noether}). The one-loop divergences preserve the
conformal invariance if  $E_{n}^{(1)}$ is finite. This important
feature is not immediately evident and has to be proved.

The conformal invariance of the divergences means that the
coefficient of the $1/\vp$ pole in $\bar{\Ga}^{(1)}_{\text{div}}$
consists of two types of terms, namely:
\ \textit{i)} the $c$-terms, i.e., functionals that
become invariant under the transformations (\ref{conftrans}) in
the limit $n=4$; and \ \textit{ii)} the $N$-terms, which may not be
invariant under these transformations but nevertheless do not contribute
to (\ref{Noether-div}). This is the case for the surface terms and
the topological terms mentioned above. Using the mapping of
divergences to the anomaly \cite{duff77}, one arrives at the
well-known classification of the terms in the anomaly
\cite{ddi,DeserSchwimmer}.

The aforementioned classification is important for understanding
the anomaly and, especially, for its integration, i.e., for the
derivation of the effective action that produces the anomaly.
This procedure has been developed in great detail for a purely
gravitational background in $2D$ \cite{polyakov81}, $4D$
\cite{rie,frts84}, and $6D$ \cite{6d} cases. The results are
known also for additional electromagnetic, scalar, and
torsion background fields in $4D$. One can therefore say that this
subject is well understood. However, all these developments are
based on the fact that the one-loop divergences are conformal, i.e.,
satisfy the conformal Noether identities (\ref{Noether-div}).
Before discussing the case of conformal quantum gravity, it is
useful to review general arguments in the semiclassical theory.

Let us analyze the arguments leading to the relation
(\ref{Noether-div}). First of all, we consider an arbitrary theory
of the fields $\phi$ with action $S[\phi, p]$, where $p$ is a set of
external parameters. In the case of gauge theory, the fields $\phi$
involve ghosts (and possible auxiliary fields) and the action 
involves gauge fixing conditions and ghost terms. The parameters 
and fields are subject to transformations $p \to p + \de p$, 
$\phi \to \phi +\delta\phi$. Here $\delta p$ are 
field-independent variations, that is $\delta \phi = X(\phi)$ with 
a local function $X(\phi)$. Our aim is to study how the effective 
action $\Ga[\phi,p]$ transforms under these transformations.

Consider the generating functional connected Green functions
$W[J,p]$ given by functional integral
\beq
e^{iW[J,p]} = \int
\mathcal{D} \phi \, e^{i(S[\phi,p] + \int \phi J)}.
\label{W}
\eeq
This leads to
\beq
\delta_{p}W[J,p] = e^{-i W [ J, p ]} \int
\mathcal{D} \phi \, ( \delta_{p}S[\phi,p] ) \,  e^{i(S[\phi,p] +
\int \phi J)}.
\label{deltapW}
\eeq
Now one fulfills replacement of
field variable in the integral $\phi \rightarrow \phi+ \delta \phi$
with $\delta \phi = X(\phi)$. This is a local replacement of
variables, the corresponding Jacobian is equal to unit within
dimensional regularization. After that we get
\beq
\delta_{p}W[J,p]
&=&
e^{-iW[J,p]}\int \mathcal{D} \phi \, \left[ \delta_{p}S[\phi,p]
+\delta_{\phi}S[\phi,p] + \int ( X(\phi)J ) \right] \,e^{i(S[\phi,p]
+ \int \phi J)}
\nn
\\
&&
\quad
=\,\,\big< \delta S[\phi,p] \big>_{J} + \int J \big< X(\phi)
\big>_{J},
\label{deltap}
\eeq
where \ $\delta S =\delta_{p} S + \delta_{\phi} S$ \
and, for any quantity $A$,
\beq
\big< A \big>_{J}
\,=\,
e^{-iW[J,p]} \int
\mathcal{D} \phi \, \, A \, \, e^{i(S[\phi,p] + \int \phi J)}.
\label{average}
\eeq
Now let us move to effective action\footnote{To
avoid cluttering the notation, we use the same letter for the mean
field and the original field $\phi$.}
\beq
\Ga[\phi,p]\,=\,W[J,p]\,-\,J\phi,
\eeq
where
\beq
\frac{\delta W[J,p]}{\delta J}=\phi
\qquad
\mbox{and}
\qquad
J=-\frac{\delta \Ga}{\delta \phi}.
\eeq
Besides, we can use the relation
\beq
\big< X(\phi) \big>_{J} = e^{-iW[J,p]}\,X
\biggl(\frac{1}{i} \frac{\delta}{\de J} \biggr)\, e^{iW[J,p]}
\,=\, X(\phi)=\delta \phi\,.
\eeq
As a result, we get the important identity
\beq
\delta {\Ga}[\phi,p]
\,=\, \big< \delta S[\phi,p] \big>_{J=-\de
\Ga / \de \phi}
\label{general-identity}
\eeq In essence, this identity is the Schwinger quantum action principle
(see, e.g., \cite{DeWitt1}).

In the conformal invariant semiclassical theory under
consideration the external metric plays the role of parameter
$p$, the set of fields $\Phi$ and the ghost fields (in gauge theories)
play the role of $\phi$,  and the role of the $X(\phi)$ is played by the
conformal transformations of all the fields. The corresponding
transformations for matter fields are given by (\ref{inftrans}). If
the classical action is conformal invariant, one might think that
the identity (\ref{general-identity}) leads to the conclusion that
the effective action is also conformal invariant. However, this is
not the case. First, quantization assumes some regularization which
can violate conformal invariance. Second, renormalization procedure
introduces the local counterterms which can also violate conformal
invariance in quantum theory. In particular, it is known that the
conformal symmetry is broken in the finite part of the renormalized
effective action \cite{duff77}. Equivalently, the anomaly
originates from the logarithmic form factors that always accompany
the UV divergences \cite{OUP}. From this perspective, the
non-invariance is caused by the presence of $\ln \Box$ and by the
non-invariance of the $\Box$ operator under (\ref{conftrans}). In
any case, this circumstance does not affect the divergent
part of $\Ga_v(g)$. One may therefore hope that the anomaly arises
precisely because the divergences are conformal.

The key point in the scheme described above is the conformal
invariance of the classical action. However, upon quantization of a
gauge vector field or the metric, the conformal invariance of the
action is broken by the Faddeev-Popov procedure, since the
gauge-fixing and ghost actions are not invariant under the local
conformal transformation (\ref{conftrans}). This means that
for the effective action divergences to be conformal invariant, the
contributions of gauges and ghosts must cancel. Such a  cancellation
was demonstrated in \cite{Adler:1976jx} for a free Abelian vector
field in $4D$, using a rather complicated approach based on
point-splitting regularization, the complete formalism of which was
presented in \cite{Chris76}. In the case of interacting theories,
including Yang-Mills fields, a complete proof within the
framework of dimensional regularization and BRST symmetry was
firstly presented in \cite{tmf}.\footnote{This cancellation
for free theory looks evident partial case of the generic result.}

In many particular cases, the conformal nature of one-loop
renormalization is known from direct calculations. However, there
exist more complicated theories in which this statement is
needed to verify whether the results of intricate calculations are
reliable. For the conformal quantum gravity theory
(\ref{actionCQG}), the conformal invariance of one-loop
renormalization is known from the direct calculations of
\cite{frts82} and \cite{antmot,Weyl}. Nevertheless, it is of utmost
importance to have a clear understanding of why this property holds.

For conformal quantum gravity coupled to quantum conformal
matter, to the best of our knowledge there exists only one work with
direct calculations \cite{BuSh86}, by two of the present authors.
The conformal invariance of the one-loop divergences was established
by means of a conformal reparametrization, as suggested earlier in 
\cite{frts82}. This reparametrization was introduced by Englert, 
Truffin, and Gastmans in \cite{ETG76}.
Later, it was discussed in the quantum context in \cite{fradvilk78}
and is sometimes referred to as conformal regularization. We discuss
this procedure in the Appendix. In any case, the results of
\cite{tmf}, together with the analysis presented in the
following sections, show that this procedure is not required.

The purpose of the present work is to address the problem described
above. First, we reconsider the proof for conformal quantum matter
given in \cite{tmf}, using a more detailed and technically simpler
consideration. We then apply this approach to prove the main
statement concerning conformal one-loop renormalization in conformal
quantum gravity, where the required cancellation between the effects
of gauge fixing and the contributions of the ghost fields is more
difficult to establish. Finally, we consider conformal quantum
gravity coupled to quantum conformal matter and prove conformal
one-loop renormalization in this case.

\section{Semiclassical theory II. The proof}
\label{sec4}

Let us discuss the one-loop renormalization of a semiclassical model
with conformal symmetry. Consider the action of a generic
GUT-like model in $n$ dimensions,
\beq
S_{0} [ \Phi,
g ] &=& \int \rd^{n} x \sqrt{-g }\, \Big\{ - \frac{1}{4}
G^{a}_{\mu \nu} G^{a \mu \nu} + i \bar{\psi}_{p} ( \ga^{\mu}
D_{\mu}^{pq} + ih^{pq}_{i} \varphi^{i} ) \psi_{q} \nn
\\
&&
+ \,\,\frac12\, g^{\mu \nu} ( D_{\mu} \varphi )^i ( D_{\nu}
\varphi )^i + \frac{1}{2} \xi_{ij} R \varphi^{i} \varphi^{j} - V (
\varphi ) \Big\}\,.
\label{actionSM}
 \eeq
 As in Sec.~\ref{sec2}, we use a general notation
 \ $\Phi = (A_{\mu}^{a}, \varphi^{i}, \psi_{p} )$ \
 for the scalar, vector and fermion fields in an external gravitational
background. The field strength of the Yang-Mills field is defined as
$G^{a}_{\mu \nu} = \pa_\mu A_\nu^a - \pa_\nu A_\mu^a
+ g f^{abc} A_\mu^b A_\nu^c$.
The covariant derivative for the scalar fields is defined as
$(D_{\mu}\ph)^i = \pa_\mu\ph^i + i g (\th^a)_{\ph}^{ij}
A_\mu^a \ph^j$, where the generators $\th^{a}_{\ph}$ are taken
in corresponding scalar representations of the gauge group. The
covariant derivative for spinor fields is similar
to scalars, with the replacement of $\pa$ by covariant derivative
of spinor in curved space and replacement of generators
$\th^{a}_{\varphi}$ by $\th^{a}_{\psi}$ in spinor representations.
The scalar potential is assumed to have the general renormalizable
quartic form,
 \beq
 V ( \varphi )
\,=\,
\frac{1}{24} \la_{i j k l} \,\varphi^{i}
\varphi^{j} \varphi^{k} \varphi^{l}.
 \eeq
Finally, $\xi_{ij}$ are nonminimal parameters of the coupling
of scalars to scalar curvature.
We consider these parameters unrestricted
for the sake of generality.

The analysis of renormalization structure is based on general
relation (\ref{general-identity}) with $S[\phi,p]$ replaced by the
total action of gauge theory,
\beq
S_{t}
\,=\, S_{0} + S_{\text{gf}} + S_{\text{gh}},
\label{acttotal}
\eeq
with $S_{0}$ from (\ref{actionSM}),
and $S_{\text{gf}},\, S_{\text{gh}}$ be the gauge fixing term and
the action of ghosts, respectively. Thus, we have the following
relation for dimensionally regularized effective action
${\Ga}[\Phi,g]$:
\beq
\de\Ga = \big< \de ( S_{0}+ S_{\text{gf}} + S_{\text{gh} } )
 \big>_{J = -\de \Ga/\de \phi }.
 \label{conformal-identity}
 \eeq
In this formula,
\beq
\delta \,=\, \int \rd^n x \sqrt{-g}\sigma(x)\mathcal{C}_n
\eeq
with the generator of conformal transformations defined in
(\ref{Noether}). Since the action $S_{0}$ is conformal invariant at
$n=4$ and $\xi_{ij}$ given by (\ref{xin}), we arrive at the
transformation rule
\beq
\delta S_{0} = \int \rd^n x
\sqrt{-g}\sigma(x) \Big\{ (n-4)E
+\Big[ \xi_{ij} - \frac{n-2}{4(n-1)} \delta_{ij} \Big] E_{ij} \Big\}.
\label{functions}
\eeq
The explicit form of the functions $E$
and $E_{ij}$ is defined by a concrete form of the action $S_{0}$.
The direct calculations in the model (\ref{actionSM}) lead to
\beq
&& E [ \Phi, g ]\,=\, -
\frac{1}{4} \, G^{a}_{\mu \nu} G^{a \mu \nu} - V ( \varphi )
+ i \bar{\psi}_{p} ( \ga^{\mu} D_{\mu}^{ pq} - h^{pq}_{i} \phi^{i})
\psi_{q},
\nn
\\
&& E^{ij} [\Phi, g ] \,=\, -2 ( n - 1 ) \na_\mu \big[(\na^{\mu}
\varphi^{i})\, \varphi^{j} \big].
 \eeq
It is worth pointing out that the explicit form of these
functions is irrelevant for further analysis.

The gauge-fixing term is taken in the form\footnote{This
gauge fixing procedure is technically
different from the  original one in \cite{tmf}. Using the gauge
fixing conditions with auxiliary field $B^\al$ slightly simplifies
the analysis in a non-Abelian theory and and is more appropriate
in the case of conformal quantum gravity.}
\beq
S_{\text{gf}} = \int \rd^{n} x \sqrt{- g}\,
\Big(B^{a} \na_{\mu} A^{\mu a} - \frac{\alpha}{2}B^{a}B^{a}\Big),
\label{gf}
\eeq
where
$B^{a}$ is the Nakanishi-Lautrup field and $\al$ is a gauge
parameter. Assuming that both $B^{a}$ and $A_{\mu}^{a}$ are inert
under conformal transformations, we find that the conformal
transformation is written as
\beq
 \mathcal{C}_n S_{\text{gf}} \,\,=\,\,- \,
 \Big[
 (n - 2) (\na_{\mu} B^{a}) A^{\mu a}
 + \frac{n\al}{2}B^{a}B^{a}\Big].
 \label{gftransf}
 \eeq
The ghost action corresponding to gauge fixing conditions
(\ref{gf}) has the form
 \beq
S_{\text{gh}} \,=\, \int \rd^{n} x \sqrt{-g} \, \bar{C}^a \na^{\mu}
D_\mu^{ a b} C^{b}\, .
\label{gh}
 \eeq
Assuming that the ghost fields do not transform under conformal
transformations, i.e.,
 \beq
 \label{eq:CTghSC}
 C^{a}\,=\, C^{*a},
\qquad \bar{C}^{a} \,=\, \bar{C}^{*a},
\label{confCC}
 \eeq
 we get
\beq \mathcal{C}_{n} S_{\text{gh}} \,\,=\,\, -\,(n - 2) \big(\na_\mu
\bar{C}^a \big) D^{\mu a b} C^b \,.
 \eeq

Now we apply the identity (\ref{general-identity}) and obtain
 \beq
 \label{eq:CnGaSC}
 \mathcal{C}_n \Ga &=&
( n - 4 ) 
\big< E ( \Phi, g ) \big>_{J} + \biggl[ \xi_{ij}\, - \,\frac{n-2}{4(
n - 1 )} \de_{ij} \biggr] \big<E^{ij} ( \Phi, g ) \big>_{J} \nn
\\
&& - \,\,(n - 2) \big< ( \na_{\mu} B^a) A^{\mu a} + ( \na^{\mu}
\bar{C}^{a} ) D_{\mu}^{ab} C^{b} \big>_{J} + \frac{n \al}{2} \big<
B^{a} B^{a}\big>_{J}.
 \label{confGam}
 \eeq
The quantum average is taken over the full set of fields
$\phi = \left( \Phi, B^{a}, \bar{C}^{a}, C^{a} \right)$ and is given
by the relation (\ref{average}) where action $S[\phi ,p]$  is $S_{t}
= S_{0} + S_{\text{gf}} + S_{\text{gh}}$ is the total action
obtained by the Faddeev--Popov quantization procedure.

The expression~\eqref{eq:CnGaSC} can be simplified by using
BRST symmetry~\cite{Becchi-1:1975nq, Becchi-2:1975nq,
Tyutin:1975qk}. Using the notation
$\Phi=(A^{a}_{\mu},\,\varphi^{i},\,\psi_{p})$,
the gauge transformations of the fields $\Phi$ can be written as
$\delta\Phi = R^{a}\xi^{a}$ with
$R^{ab}=(D^{ab}_{\mu},
\,i(\th_{\varphi}^{a})^{ij}\varphi^{j},\,i(\th_{\phi}^{a})
_{pq}\psi_{q})$. Then, the BRST transformations which leave the
action $S_{t}[\Phi, B, \bar{C}, C]$ invariant are written as follows
\beq
&&
\de_{\text{BRST}}\Phi \,=\, R^{a} C^{a} \la, \nn
\\
\nn
&&
\de_{\text{BRST}} B^{a} \,=\, 0,
\nn
\\
&&
\de_{\text{BRST}} \bar{C}^{a} \,=\, B^{a} \la,
\nn
\\
&&
\de_{\text{BRST}} C^{a} \,=\,\frac{1}{2}\, f^{abc}\, C^b C^c \la,
\label{BRST}
\eeq
where $\la$ is a  global anticommuting parameter.

Let $F[\phi]$ be an arbitrary functional. Its average value
is defined according to (\ref{average}),
\beq
\big< F[\phi] \big>_{J}
&=&
e^{-iW[J]} \int \mathcal{D}
\phi \,\, F[\phi] \,\,e^{i(S_{t} +\int \rd^{n}x \sqrt{-g} \phi J)},
 \label{average1}
 \eeq
where the sources are introduced only to the fields $\Phi$. Making a
substitution of variables corresponding to BRST transformations
(\ref{BRST}) and taking into account that the Jacobian is equal to
unit, one gets
\beq
\big< F \big>_{J} =  \big< F \big>_{J}  + \big< \de_{\text{BRST}}
F + i\int \rd^{n}x \sqrt{-g} \de_{\text{BRST}}\phi J \big>_{J}\,.
\eeq
This relation leads to the general identity
\beq
\big< \de_{\text{BRST}} F + i\int \rd^{n}x \sqrt{-g}
\,F \de_{\text{BRST}}\,\phi \,J\big>_{J} \,=\, 0,
\label{identityBRST}
 \eeq
where  $\phi = (\Phi,B,\bar{C},C)$,  $\Phi = (A, \varphi, \psi)$,
and we took into account $\la^{2}=0$ and
$\de_{\text{BRST}}S_{t}=0$.

Consider the identity (\ref{identityBRST}) with $F$ replaced
by the function $\bar{C}^{a}(x)A^{a}_{\mu}(x^{\prime})$.
The BRST-transformation of this expression has the form
\beq
\de_{\text{BRST}} \bar{C}^{a}(x)A^{a}_{\mu}(x^{\prime})
\,=\,
\big[
B^{a}(x)A^{a}_{\mu}(x^{\prime})
+ \bar{C}^{a}(x)D^{ab}_{\mu}C^{b}(x^{\prime})
\big]\la.
\eeq
Substituting this result to Eq.~(\ref{identityBRST}) and writing
the BRST transformations as
\beq
\de_{\text{BRST}} \Phi \,=\, R_{b} ( \Phi )\, C^{b} \la,
\eeq
where $R_{b} ( \Phi )$ are the corresponding generators, one obtains the identity
 \beq
&&
\big< B^{a}(x)A^{a}_{\mu}(x^{\prime})
+ \bar{C}^{a}(x)D^{ab}_{\mu}C^{b}(x^{\prime})\big>_{J} \la
\nn
\\
&&
\quad
+\,
i\int \rd^{n}y \sqrt{-g} \, \big< \bar{C}^{a}(x) A^{a}_{\mu}( x^{\prime} )
R_{b} ( \Phi ( y ) ) C^{b} ( y ) \big>_{J} \la J (y)=0.
 \label{identity1}
  \eeq
Taking covariant derivative with respect $x$ and putting
$x^{\prime}=x$, one gets
\beq
&&
\big< ( \nabla^{\mu}B^{a} ) A^{a}_{\mu}
+ ( \nabla^{\mu}\bar{C}^{a} ) D^{ab}_{\mu}C^{b}\big>_{J}
\nn
\\
&&
\quad
= \,-\,i ( -1 )^{\varepsilon ( \Phi )} \int \rd^{n}y
\sqrt{-g} \, \big< ( \nabla^{\mu}\bar{C}^{a}(x) ) A^{a}_{\mu}(x)
R_{b} ( \Phi ( y ) ) C^{b} ( y ) \big>_{J}J(y).
\label{identity2}
\eeq
Analogously, using $\bar{C}^{a}(x)B^{a}(x^{\prime})$ as
$F$ in the identity (\ref{identityBRST}) and setting $x^{\prime}=x$,
we get
\beq
\label{identity3}
\big< B^{a} B^{a} \big>_{J}
\,=\,
i ( -1 )^{\varepsilon (\Phi) + 1} \int \rd^{n} y
\sqrt{-g} \, \big< \bar{C}^{a} ( x ) B^{a} ( x )
R_{b} ( \Phi ( y ) ) C^{b} ( y ) \big>_{J} J ( y ),
\eeq
where $\varepsilon ( \Phi )$ denotes the Grassmann parity
of the field $\Phi$.

Let us now turn to identity (\ref{confGam}) and use
(\ref{identity2}) and (\ref{identity3}). As a result, we obtain
\beq
 \mathcal{C}_n \Ga &=&
(n - 4)  \big< E ( \Phi, g ) \big>_{J}
+ \Big[ \xi_{ij}\,-\,\frac{n-2}{4( n - 1 )} \de_{ij} \Big]
\big<E^{ij} ( \Phi, g ) \big>_{J}
\nn
\\
&&
+\, (-1)^{\varepsilon ( \Phi )} \,i\,(n - 2) \int \rd^{n}y \sqrt{-g}
\,\big< (\nabla^{\mu}\bar{C}^{a}(x) ) A^{a}_{\mu}(x)
R_{b} ( \Phi ( y ) ) C^{b} ( y ) \big>_{J}J(y)
\nn
\\
&&
+\,(-1)^{\varepsilon (\Phi) + 1}\,
\frac{ n \al i}{2}  \int \rd^{n} y \sqrt{-g}
\big< \bar{C}^a(x) B^a (x) R_b(\Phi(y))C^b(y) \big>_{J} J ( y ).
\label{eq:CnGaSC-1}
\eeq
The above relation is the final form of the conformal Ward identity.
Although it is valid to all orders in perturbation theory, its practical
applications are restricted to the one-loop approximation. To see
this, let us analyze the last two terms appearing in
Eq.~\eqref{eq:CnGaSC-1}. Performing the loop expansion,
\beq
\label{eq:loopexp}
\Phi \to \tilde{\Phi} + \hbar^{1/2} \Phi,
\quad
B^{a} \to \hbar^{1/2} B^{a},
\quad
\bar{C}^{a} \to \hbar^{1/2} \bar{C}^{a},
\quad
C^{a} \to \hbar^{1/2} C^{a},
\eeq
with $\tilde{\Phi}$ denoting the background matter fields, one
finds that the third term in~\eqref{eq:CnGaSC-1} is of the order
$\mathcal{O}(\hbar^{3/2})$. Then the second term in
(\ref{eq:CnGaSC-1}) gets reduced to
\beq
\label{eq:secterm}
-\,i (n-2) \,
\hbar \tilde{A}_{\mu}^{a} ( x )  \na^{\mu}_{x}
\int \rd^{n} y \sqrt{-g} \, R_{b} ( \tilde{\Phi} ( y ) )
\big< \bar{C}^{a} ( x ) C^{b} ( y ) \big> J ( y )
\,+\, \mathcal{O} ( \hbar^{3/2} ),
\eeq
where we took into account that $\vp( R_{b} ) = \vp(\Phi)+1$.
In the one-loop approximation,
the quantum average is taken with the quadratic action
\beq
\big< F[\phi] \big>
&=&
\frac{\int \mathcal{D}
\phi \,F[\phi] \, e^{\frac{i}{2} S_{t}^{(2)} \phi^2}}{\int \mathcal{D}
\phi \,e^{\frac{i}{2} S_{t}^{(2)} \phi^2}}.
\label{average2}
\eeq
However, Eq.~\eqref{eq:secterm} is already of the one-loop
order. Therefore, within our approximation, one can write the
source term as
\beq
J (y) \,=\, -\,\frac{\de S_{t}}{\de \Phi ( y )},
\eeq
which is a finite quantity. The resulting expression in
Eq.~\eqref{eq:secterm} then contains only a single ghost
propagator attached to the source. Consequently, it cannot
produce one-loop ultraviolet divergences. As a result, neither
the second nor the third term in~\eqref{eq:CnGaSC-1} contributes
to the divergent part of the one-loop effective action. On top of
that, if we choose $\xi_{ij}$ according to (\ref{xin})
and take the limit $n \to 4$, we finally arrive at
\beq
\mathcal{C}_4 \,\bar{\Ga}^{(1)}_{\text{div}} \,=\, 0\,,
\eeq
which is precisely this result that was obtained in \cite{tmf}.

\section{Conformal quantum gravity}
\label{sec5}

Let us consider the proof of one-loop conformal renormalizability
for the conformal quantum gravity. The classical action of the theory
takes the form equivalent to (\ref{actionCQG})
\beq
\label{eq:action}
S_{0} = \int \rd^n x \sqrt{- g }\,\mathscr{L}_0 ( g )
\,,\qquad
\mathscr{L}_{0} ( g )
= -\frac{1}{2 \la} C^2 + \frac{1}{\rh} E_{4} + \ta \Box R.
\eeq
The Gauss-Bonnet term, as well as the term $\Box R$, are included
for the sake of multiplicative renormalizability. Correspondingly,
$\rho$ and $\tau$ are free parameters. These parameters are subjects
to the renormalization procedure, but the corresponding terms do
not affect the quantum corrections. It is worth mentioning that this
issue has been explored in detail \cite{Weyl}, following the original
proposal of \cite{capkim}.

A more convenient approach to studying one-loop divergences
involves the use of the background field method. Thus, we split
the metric according to
\beq
g_{\mu \nu} \,\longrightarrow \,
g_{\mu \nu}^{\prime} = g_{\mu \nu} + h_{\mu \nu},
\eeq
where $g_{\mu \nu}$ denotes the background metric and $h_{\mu \nu}$
is the quantum field to be integrated over in the functional integral.
The next step is to perform the Faddeev-Popov procedure. This
method, in the context of higher-derivative gravity models, requires
a special adaptation introduced in \cite{frts82}. It has recently
been described in detail, for example, in the book~\cite{OUP} and in
the review papers \cite{Ohta:2020bgz,Handbook-QG,Lavrov:2022ceg}.

The gauge-fixing and ghost actions can be introduced in different,
albeit equivalent, ways, some of which are more convenient for
practical one-loop calculations, while others are more suitable for
proving general statements. In what follows, we adopt a
choice that leads to a particularly simple form of the BRST
transformations. Thus, the gauge-fixing action is taken in the form
\beq
 S_{\text{gf}} = \int \rd^{n} x \sqrt{- g }
 \, B_{\al} Y^{\al \be} \chi_{\be},
\label{actgf}
\eeq
where $B^{\al}$ is the Nakanishi-Lautrup field,
which is a vector in case of quantum gravity. Furthermore,
$\chi^{\al}$ is the gauge-fixing condition for diffeomorphism
invariance,
\beq
\label{eq:GFcond}
\chi_{\al} = \na_{\la} h^{\la}_{\al} + \be \na_{\al} h.
\eeq
In this formula, $h = g^{\mu \nu} h_{\mu \nu}$ is the trace of the
quantum metric field. To fix the additional conformal symmetry, one
may impose the condition $h = 0$~\cite{frts82}, which implies
\beq
\label{eq:GFconf}
1 + n \be = 0.
\eeq
However, we will keep $\be$ arbitrary until the end of our
considerations. The next element of (\ref{actgf}) is the weight
operator introduced in \cite{frts82},
\beq
Y^{\al \be} = g^{\al \be} \Box + \ga \na^{\al} \na^{\be}.
\label{weight}
\eeq
One can add additional terms with arbitrary gauge parameters to this
formula. However, this would not make much sense, since we know
that the one-loop divergences in the conformal model
\eqref{eq:action} are independent of the gauge-fixing
parameters~\cite{frts82,a}.

Finally, the corresponding ghost action is given by
\beq
\label{eq:Sgh}
S_{\text{gh}}
= \int \rd^{n} x \sqrt{- g } \, \bar{C}_\al \,Y^{\al}{}_{\ga}
M^{\ga}{}_{\be} C^{\be}
\,-\, \frac{\al_{2}}{2}  \int \rd^n x \sqrt{- g }
\, B_{\al} Y^{\al \be} B_{\be},
\eeq
where $C^{\al}$ and $\bar{C}^{\al}$ are the Grassmann-odd
Faddeev-Popov ghost and antighost fields, respectively, and
$\al_{2}$ is an arbitrary parameter.
The operator $M^{\al}{}_{\be}$ is defined by
\beq
M^{\al}{}_{\be}
\,=\, \frac{\de \chi^{\al}}{\de h_{\mu \nu}} R_{\mu \nu, \be}
\,=\, \de^\al_{\,\be} \Box + \na^\be \na^\al
+ 2 \be \na^\al \na_\be,
\eeq
where $R_{\mu \nu, \al}$ is the gauge transformations generator
for the field $h_{\mu \nu}$ \cite{Stelle77}. With the choice of the
weight operator $Y^{\al\be}$ given in (\ref{weight}), the ghost
operator is fourth order in derivatives. The total action after the
Faddeev-Popov procedure has the form
\beq
\label{eq:St}
S_t \,=\, S_{0} \,+\, S_{\text{gf}} + S_{\text{gh}}.
\eeq

Consider an infinitesimal conformal transformation (\ref{inftrans})
for the metric,
\beq
\label{eq:CTh}
g'_{\mu \nu} 
= 2 \si \bar{g}'_{\mu \nu} 
= 2 \si ( \bar{g}_{\mu \nu} + \bar{h}_{\mu \nu} ),
\eeq
where barred quantities have indices raised and lowered using
the metric $\bar{g}_{\mu \nu}$. The conformal transformation
properties of the relevant quantities can be found, for example,
in~\cite{Stud}. The classical action transforms as
\beq
\mathcal{C}_n S_{0} ( g^{\prime} )
\,=\, (n - 4) \,\mathscr{L}_{0} ( g^{\prime} ).
\eeq

The definition of the conformal transformation for the extra
fields in the total action (\ref{eq:St}) has to be made on the
basis of convenience.
The ghost and antighost fields, as well as the Nakanishi-Lautrup
field $B_{\al}$, do not transform under conformal transformations,
\beq
\label{eq:CTB}
\bar{C}_{\al} = \bar{C}_{\al}^{*},
\qquad
C_{\al} = C_{\al}^{*},
\qquad
B_{\al} = \bar{B}_{\al}.
\eeq

From Eq.~\eqref{eq:GFcond}, the gauge-fixing condition transforms as
\beq
\chi_{\al} = \bar{\chi}_{\al}
+ n \bar{h}_{\al \be} \bar{\na}^{\be} \si
- \bar{h} \bar{\na}_{\al} \si.
\eeq
The gauge-fixing action transforms according to (\ref{eq:CTh}),
\beq
\mathcal{C}_n S_{\text{gf}}
\,=\, (n - 4) \mathscr{L}_{\text{gf}}
\,+\, \na_{\al} E^{\al}_{\text{gf}}
\,+\, \na_\al \na_\be F^{\al \be}_{\text{gf}},
\eeq
where
\beq
\mathscr{L}_{\text{gf}}
&=& - \,(\na_{\al} \chi_{\be} ) ( \na^{\al} B^{\be} )
- \ga ( \na_{\al} \chi^{\al} ) ( \na_{\be} B^{\be} ),
\nn
\\
E^{\al}_{\text{gf}}
&=&
-\,\na^{\al} ( \chi_{\be} B^{\be} ) - \chi^{\be} ( \na_{\be} B^{\al} )
- ( \na_{\be} \chi^{\al} ) B^{\be} - ( \na_{\be} h ) ( \na^{\be} B^{\al} )
\nn
\\
&&
+ \, n \ga ( \na_{\be} h^{\al \be} ) ( \na_{\ga} B^{\ga} )
+ n ( \na_{\ga} h^{\al}_{\be} ) ( \na^{\ga} B^{\be} )
- \ga ( \na^{\al} h ) ( \na_{\be} B^{\be} )\nn
\\
&&
+ \, \big(1 + n\ga - 2\ga \big)
\big[\chi^\al (\na_\be B^\be) + (\na_\be \chi^\be) B^{\al} \big],
\nn
\\
F_{\text{gf}}^{\al \be}
&=&
h ( \nabla^{\be} B^{\al} ) - n h^{\al}_{\ga} ( \na^{\be} B^{\ga} )
- n \ga h^{\al \be} ( \na_{\ga} B^{\ga} )
+ \ga h g^{\al \be} ( \na_{\ga} B^{\ga} ).
\label{contransgf}
\eeq

In the ghost sector, after integrating by parts, we recast the ghost
action~\eqref{eq:Sgh} as
\beq
S_{\text{gh}}
&=&
\int \rd^{n} x \sqrt{- g }\,
( Y_{\ga}{}^{\al} \bar{C}_{\al} ) M^{\ga}{}_{\be} C^{\be}
\nn
\\
&&
+ \,\frac{\al_{2}}{2} \int \rd^{n} x \sqrt{- g }
\left\{ ( \na_{\be} B_{\al} ) ( \na^{\be} B^{\al} )
+ \ga ( \na_{\al} B^{\al} ) ( \na_{\be} B^{\be} ) \right\}.
\eeq
In this way, we avoid the need to compute the conformal
transformation of a fourth-order differential operator. The
resulting conformal variation takes the form
\beq
\mathcal{C}_n S_{\text{gh}}
\,=\, (n - 6) \mathscr{L}_{\text{gh}}
\,+\, \na_{\al} E^{\al}_{\text{gh}}
\,+\, \na_{\al} \na_{\be} F^{\al \be}_{\text{gh}},
\label{contransgh}
\eeq
where
\beq
\mathscr{L}_{\text{gh}}
&=&
- \, \big( M^{\ga}{}_{\be} C^{\be} \big)
\big( Y_{\ga}{}^{\al} \bar{C}_{\al} \big)
\,+\, \frac{\al_{2} ( n - 4 )}{2 ( n - 6)}
\left[ ( \na_{\al} B_{\be} )^2 + \ga ( \na_{\al} B^{\al} )^2 \right] ,
\nn
\\
E^{\al}_{\text{gh}}
&=&
(n - 4) \big[(\na^{\al} C^{\be} ) ( Y_{\be \ga} \bar{C}^{\ga} )
+ ( M^{\be \ga} C_{\ga} ) (\na^{\al} \bar{C}_{\be} ) \big]
- 2 ( 1 + \ga ) ( M^{\al}{}_{\be} C^{\be} ) ( \na_{\ga} \bar{C}^{\ga} )
\nn
\\
&&
- \,4 ( 1 + \be ) ( \na_{\be} C^{\be} ) ( Y^{\al}{}_{\ga} \bar{C}^{\ga} ) + [ 2 (n-2) \be + n ] ( \na^{\be} C^{\al} ) ( Y_{\be \ga} \bar{C}^{\ga} )
    \nn
    \\
&& + \,
(2 + n\ga - 2\ga) \big( M_{\be \ga} C^\ga\big)
\big(\na^{\be} \bar{C}^{\al}\big)
+ \al_{2} B^{\be} \left[ ( \na^{\al} B_{\be} )
+ \big( \na_{\be} B^{\al} \big) \right]
\nn
\\
&&
- \,\al_{2} \left( 1 +  n\ga -2\ga \right) B^\al
\big(\na_{\be} B^{\be} \big) \,,
\nn
\\
F_{\text{gh}}^{\al \be}
&=&
2 g^{\al \be} C_{\ga} \big( Y^{\ga \de} \bar{C}_{\de} \big)
+ 2 C^{\be} ( Y^{\al}{}_{\ga} \bar{C}^{\ga} )
+ g^{\al \be} ( M_{\ga \de} C^{\de} ) \bar{C}^{\ga}
+ ( M^{\al}{}_{\ga} C^{\ga} ) \bar{C}^{\be}
\nn
\\
&&
-\, 2 (1 + n\be - 2\be) C^{\al} ( Y^{\be}{}_{\ga} \bar{C}^{\ga} )
- (1 + n\ga - 2\ga) \big( M^{\be}{}_{\ga} C^\ga \big) \bar{C}^\al.
\eeq
It is useful to define
\beq
\mathscr{L}
&=&
\mathscr{L}_{0} + \mathscr{L}_{\text{gf}} + \mathscr{L}_{\text{gh}},
\nn
\\
E^{\al}
&=&
E^{\al}_{\text{gf}} + E^{\al}_{\text{gh}},
\nn
\\
F^{\al \be} &=& F^{\al \be}_{\text{gf}} + F^{\al \be}_{\text{gh}}\,.
\eeq
Upon the standard averaging procedure, the total result for the
effective action is
\beq
\label{eq:CnGa0beforeBRST}
\mathcal{C}_{n} \Ga_{}
\,=\,
(n - 4) \big< \mathscr{L} ( g^{\prime} ) \big>
- 2 \big< \mathscr{L}_{\text{gh}} \big>
+ \big< \na_{\al} E^{\al} \big>
+ \big< \na_{\al} \na_{\be} F^{\al \be} \big>\,.
\eeq

To analyze this expression, we use the invariance of the
total action (\ref{eq:St}) under gravitational BRST
transformations~\cite{Stelle77} (see, e.g.,  \cite{Lavrov:2022ceg}
for details of the BRST transformation in the case of gauge with
the Nakanishi-Lautrup field)
\beq
\de_{\text{BRST}} h_{\mu \nu} &=& ( \na_{\mu} C_{\nu}
+ \na_{\nu} C_{\mu} ) \la,
\nn
\\
\de_{\text{BRST}} B^{\al} &=& 0,
\nn
\\
\de_{\text{BRST}} C^{\al} &=& -C^{\be} \pa_{\be} C^{\al} \la,
\nn
\\
\de_{\text{BRST}} \bar{C}^{\al} &=& B^{\al} \la,
\label{eq:BRSTgrav}
\eeq
where $\la$ is a constant Grassmann-odd parameter. As in the
Yang–Mills theory described in the previous section, the BRST
identities provide relations between the propagators and,
ultimately, explain why the one-loop divergences are conformal.

Taking $F = h_{\mu \nu} ( x ) \bar{C}_{\al} ( x^{\prime} )$ in the
identity~\eqref{identityBRST} and setting the sources to zero gives
\beq
\label{eq:hb}
\big< h_{\mu \nu} B_{\al} \big>
\,=\, \big< ( \na_{\mu} C_{\nu} ) \bar{C}_{\al} \big>
\,+\, \big< ( \na_{\nu} C_{\mu} ) \bar{C}_{\al} \big>\,.
\eeq
This relation is important, as it enables one to express the mixed
propagator in terms of the ghost propagators. Using the definition
of $\chi_{\al}$, one readily obtains
\beq
\big< \chi_{\al} B_{\be} \big>
\,=\, \big< ( \na_{\la} h^{\la}_{\al} ) B_{\be} \big>
\,-\, \be \big< ( \na_{\al} h ) B_{\be} \big>\,.
\label{form1}
\eeq
Taking a covariant derivative of  Eq.~(\ref{eq:hb})
with respect to the argument of $h_{\mu \nu}$, we get
\beq
\big< ( \na_{\rh} h_{\mu \nu} ) B_{\al} \big>
\,=\,
\big< ( \na_{\rh} \na_{\mu} C_{\nu} ) \bar{C}_{\al} \big>
\, + \,
\big< ( \na_{\rh} \na_{\nu} C_{\mu} ) \bar{C}_{\al} \big>\,.
\label{form2}
\eeq
Replacing (\ref{form2}) in (\ref{form1}) gives
\beq
\big< \chi_{\al} B_{\be} \big>
\,=\,
\big< ( \Box C_{\al} ) \bar{C}_{\be} \big>
+ \big< \na_{\la} \na_{\al} C^{\la} ) \bar{C}_{\be} \big>
- 2 \be \big< \na_{\al} \na_{\mu} C^{\mu} ) \bar{C}_{\be} \big>.
\label{form3}
\eeq
Thus, we arrive at the useful identity
\beq
\big< \chi_{\al} B_{\be} \big>
\,=\, \big< M_{\al}{}^{\ga} C_{\ga}  \bar{C}_{\be} \big>.
\label{usefulid}
\eeq
derived from the gauge-fixing condition $\chi_{\al}$ and
Eq.~\eqref{eq:hb}.

Furthermore, taking
$F = \bar{C}_\al ( x ) B_\be ( x^{\prime} )$
in~\eqref{identityBRST} and switching off the sources, we obtain
\beq
\label{eq:bb}
\big< B_{\al} (x) B_{\be} (x^\prime) \big> = 0
,
\eeq
which gives, in the coincidence limit $x \to x^\prime$,
\beq
\big<  B_{\al} B_{\be} \big> \,=\, 0\,.
\eeq
Thus, the contribution of the second term in the
action~\eqref{eq:Sgh} is irrelevant for the conformal variation
of the effective action $\mathcal{C}_{n} ( x ) \Ga$.

At this point, a comment is in order. Unlike the semiclassical theory
described in the previous section, the above identities cannot
guarantee a complete cancellation between the contributions of the
gauge-fixing and ghost sectors. Eq.~\eqref{eq:hb} already involves
a derivative acting on the ghost field, unlike the conformal
identities in the semiclassical case. On the other hand, the
conformal variation of the ghost action produces terms in which
all four derivatives act on the antighost field $\bar{C}^{\alpha}$.
Such terms cannot be cancelled by contributions arising from the
gauge-fixing sector through the identity~\eqref{eq:hb}. Thus, we
need to analyze the remaining terms in detail.

Replacing Eqs.~\eqref{eq:hb} and \eqref{eq:bb} into
Eq.~\eqref{eq:CnGa0beforeBRST} is a computation that can
be conveniently carried out using
\textit{Mathematica}~\cite{Mathematica} and the package
\texttt{FieldsX}~\cite{Frob:2020gdh}. Thus, we arrive at
\beq
\mathcal{C}_{n} \Ga
&=&
(n - 4) \big< \mathscr{L}_{0} ( g^{\prime} ) \big>
\,+\, 2 \big< C_\al M^\al{}_\ga Y^\ga{}_\be \bar{C}^\be \big>
\nn
\\
&&
\,-\, 2 (1 + n\be) \big< \na_\al(C^\al \na_\ga Y^\ga{}_{\be}
\bar{C}^\be)\big>.
\label{mainid}
\eeq
This is a remarkable result. The first term in this expression does
not contribute to the divergent part of the one-loop effective action
in the limit $n \to 4$ because the classical action is conformally
invariant.\footnote{In contrast, for a nonconformal theory,
$C_n(x) S_0 \neq 0$, so the result is not proportional to $(n-4)$.}
Next, the third term in (\ref{mainid}) vanishes because of
the gauge-fixing condition $h=0$, which is equivalent to
condition \eqref{eq:GFconf}.

It remains to analyze the contribution of the second term in
(\ref{mainid}). To do this, we use the fact that the functional
integral of a total derivative vanishes. In particular, we meet 
\beq
\label{eq:DSeq}
\int \mathcal{D} \phi \,\frac{\de_{r}}{\de C^{\be} (x^{\prime})}
\,e^{i \left( S_{t} [ \phi, g^{\prime} ] + \phi \mathcal{J} \right)}
\,=\,
\int \mathcal{D} \phi \Big[
i \frac{\de_{r} S}{\de C^{\be} ( x^{\prime} )}
+ i \bar{J}_{\be} ( x^{\prime} ) \Big]
e^{i  \left( S_{t} [ \phi, g^{\prime} ] + \phi \mathcal{J} \right)}
\,=\, 0\,,
\eeq
where $\de_{r}/\de$
is the right functional derivative and we introduced
\beq
\phi = \left( h_{\mu \nu}, B^{\al}, C^{\al}, \bar{C}^{\al} \right)
\quad
\mbox{and}
\quad
\mathcal{J}
= \big( J^{\mu \nu}, J_{\al}^{(B)},\bar{J}_{\al}, J_{\al}\big),
\label{PsiJ}
\eeq
as collective notations for all the fields and respective
sources. The condensed form is
\beq
\phi \mathcal{J} = \int \rd^{n} x \sqrt{- g }
\left\{ h_{\mu \nu} J^{\mu \nu} + B^{\al} J_{\al}^{(B)}
+ \bar{J}_{a} C^{a} + \bar{C}^{a} J_{a} \right\}.
\eeq
Taking the right functional derivative of~\eqref{eq:DSeq} with
respect to $\bar{J}_{\al} ( x )$, we obtain
\beq
\int \mathcal{D} \phi \left\{ i \de^{\al}_{\be} \de ( x - x^{\prime} )
+ \frac{\de_{r} S_{t}}{\de C^{\be} ( x^{\prime} )} C^{\al} ( x )
+ \bar{J}_{\be} ( x^{\prime} ) C^{\al} ( x ) \right\}
e^{i \left( S_{t} [ \phi, g^{\prime} ] + \phi \mathcal{J} \right)}\,=\,0\,.
\eeq
Setting the sources to zero, we arrive at
\beq
\Big< C^{\al} (x) \,\frac{\de_{r} S}{\de C^{\be} (x^{\prime})}\Big>
\,=\, i \de^{\al}_{\,\be}\, \de (x - x^{\prime}).
\eeq
Using the fact that
\beq
\frac{\de_{r} S}{\de C_{\be} ( x^{\prime} )}
\,=\,
M^{\be}{}_{\ga} Y^{\ga}{}_{\de} \bar{C}^{\de} (x^{\prime}),
\eeq
that follows from~\eqref{eq:Sgh}, we obtain
\beq
\big< C_{\al} ( x ) M^{\be}{}_{\ga}
Y^{\ga}{}_{\de} \bar{C}^{\de} ( x^{\prime} ) \big>
\,=\, i \de^{\be}_{\,\al}\, \de (x - x^\prime).
\eeq
Contracting the indices $\al$ and $\be$ and taking the coincidence
limit $x^{\prime} \to x$, the right-hand side becomes proportional
to $\delta ( 0 )$, which vanishes in dimensional
regularization (see, e.g., ~\cite{Leibbrandt:1975dj}). Therefore,
\beq
\big< C_\al M^\al{}_\ga Y^{\ga}{}_{\be}\bar{C}^\be\big> = 0\,.
\eeq
In total, the final result is simply
\beq
\label{eq:finalCQG}
\mathcal{C}_{n} \Ga
\,=\, (n - 4) \big< \mathscr{L}_{0} ( g^{\prime} ) \big>.
\label{eq:CnGaCQGfinal}
\eeq
Separating the one-loop effective action $\Ga$ into finite and
divergent $\mathcal{O}\big((n-4)^{-1}\big)$ parts and taking
the limit $n \to 4$, we obtain
\beq
\label{eq:C4G_CQG}
\mathcal{C}_4 \,\bar{\Ga}_{\text{div}}^{(1)} \,=\, 0,
\eeq
that is the relation we intended to prove in the case of pure
conformal quantum gravity.

Power-counting arguments show that the logarithmic divergences
in this theory contain four derivatives of the metric. This
number corresponds to the possible counterterms of the types
$C^2$, $E_4$, $\Box R$, and $R^2$.
The relations (\ref{eq:CnGaCQGfinal}) and
(\ref{eq:C4G_CQG}) demonstrate that the $R^2$ term cannot
appear in the one-loop divergences because it contradicts
(\ref{eq:C4G_CQG}). This is also consistent with complicated
direct calculations \cite{antmot}, independently confirmed in
\cite{Weyl}.

It is worth noting that the $R^2$ term is present in the finite part
of the effective action as a result of integrating the $\Box R$-term
in the anomaly. Obviously, this does not contradict
(\ref{eq:C4G_CQG}). Concerning the divergences, their general
structure at the one-loop level is
\beq
\bar{\Ga}^{(1)}_{\text{div}} \,=\,
-\,\frac{\mu^{n-4}}{\ep} \int \rd^{n} x \sqrt{- g }
\left\{ k_{1} C^2 + k_{2} E_{4} + k_{3} \Box R
\right\},
\label{Gadiv}
\eeq
where $\ep = ( 4 \pi )^2 ( n - 4 )$. In the semiclassical theory, the
$\Box R$ term is subject to the well-known ambiguity \cite{duff94}. 
In dimensional and Pauli-Villars regularizations,
this ambiguity is equivalent to the freedom to modify the classical
vacuum action \cite{anomaly-2004}. In conformal quantum gravity, 
as argued above, there is no such freedom, but the ambiguity persists.

\section{Conformal quantum gravity with matter}
\label{sec6}

In this section, we demonstrate the one-loop renormalizability of
conformal quantum gravity coupled to conformal matter fields.
The classical action is given by unification of the two previously
considered actions (\ref{actionSM}) and (\ref{eq:action}),
\beq
\label{eq:CQGwMF}
S_{0} [ \Phi, g ]
&=&
\int \rd^{n} x \sqrt{- g } \biggl\{
- \frac{1}{2 \la} C^2
+ \frac{1}{\rh} E_{4}
+ \ta \Box R
- \frac{1}{4} G^{a}_{\mu \nu} G^{a \mu \nu}
+ \frac{1}{2} g^{\mu \nu} (D_{\mu} \phi)^i (D_{\nu} \phi )^i
\nn
 \\
&&
+ \,\,\frac{1}{2} \xi_{ij} R \phi^{i} \phi^{j}
- V ( \phi )
+ i \bar{\psi}_{p} ( \ga^{\mu} D_{\mu}^{pq}
- h^{pq}_{i} \phi^{i} ) \psi_{q} \biggr\}.
\eeq
As before, we apply the background field method
\beq
&&
\phi^{i} \, \longrightarrow \, \phi^{\prime i} = \phi^{i} + \ph^i,
\qquad
A_{\mu}^{a}  \,\longrightarrow \, A^{\prime a}_{\mu}
= \mathcal{B}^{a}_{\mu} + A^{a}_{\mu},
\nn
\\
&&
\psi_{p} \, \longrightarrow \, \psi_{p}^{\prime}
= \psi_{p} + \eta_{p},
\qquad
g_{\mu \nu} \, \longrightarrow \, g_{\mu \nu}^{\prime}
= g_{\mu \nu} + h_{\mu \nu}.
\eeq
The gauge-fixing procedure must simultaneously account for
Yang-Mills gauge symmetry, diffeomorphism invariance, and
conformal symmetry. This is achieved by introducing the
gauge-fixing action
\beq
S_{\text{gf}} = \int \rd^{n} x \sqrt{-g}
\, B^{a} \na_{\mu} A^{\mu a}
+ \int \rd^{n} x \sqrt{-g}
\,B_{\al} Y^{\al \be} \chi_{\be},
\label{gf-1}
\eeq
supplemented by the conformal gauge condition~\eqref{eq:GFconf}.

The corresponding ghost action is
\beq
S_{\text{gh}} \,&=&\,
\int \rd^{n} x \sqrt{-g}
\, \bar{C}^a \na^{\mu} D_\mu^{ a b} C^{b}
\,+\,
 \int \rd^{n} x \sqrt{- g } \, \bar{C}_\al \,Y^{\al}{}_{\ga}
M^{\ga}{}_{\be} C^{\be}
\nn
\\
\,&+&\,
\frac{\al_{1}}{2} \int \rd^{n} x \sqrt{-g} \, B^a B^a\,
\,-\, \frac{\al_{2}}{2}  \int \rd^n x \sqrt{- g }
\, B_{\al} Y^{\al \be} B_{\be}.
\label{gh-1}
\eeq

The conformal variation of the effective action can be derived
without additional efforts by combining Eqs.~\eqref{eq:CnGaSC-1}
and~\eqref{eq:CnGa0beforeBRST}. However, in the previous section
we considered only the vacuum part of the effective action. In the
present case, we have to introduce external sources. Using the
identity~\eqref{identityBRST} it is possible to find the modified
versions of the relations~\eqref{eq:hb} and~\eqref{eq:bb}, which
have the form
\beq
&&
\big< h_{\mu \nu} ( x ) B_{\al} ( x^{\prime} )
- ( \na_{\mu} C_{\nu} ( x )
+ \na_{\nu} C_{\mu} ( x ) ) \bar{C}_{\al} ( x^{\prime} ) \big>_{J}
\nn
\\
&&
\qquad \qquad
= \, -(-1)^{\varepsilon ( \Phi )} i\int \rd^{n} y \sqrt{-g}
\, \big< h_{\mu \nu} ( x ) \bar{C}_{\al} ( x^{\prime} )
R_{b} ( \Phi ( y ) ) C^{b} ( y ) \big>_{J} J ( y )
\eeq
and
\beq
\big< B_{\al} ( x ) B_{\be} ( x^{\prime} ) \big>_{J}
= -(-1)^{\varepsilon ( \Phi )} i \int \rd^{n} y \sqrt{-g}
\, \big< \bar{C}_{\al} ( x ) B_{\be} ( x^{\prime} )
R_{b} ( \Phi ( y ) ) C^{b} ( y ) \big>_{J} J ( y ).
\eeq
Performing a loop expansion similar to~\eqref{eq:loopexp}, we see
that the right-hand sides of the above identities are already of
order $\mathcal{O}(\hbar^{3/2})$, so that they do not contribute
to the one-loop effective action and will therefore be neglected.
As a result, the conformal variation of the effective action is given by
\beq
&&
C_{n} \Ga \,=\,
( n - 4 ) [ \big< \mathscr{L}_{0} ( g^{\prime} ) \big>_{J}
+ \big< F ( \Phi^{\prime}, g^{\prime} ) \big>_{J} ]
\nn
\\
&&
\qquad \quad
+ \,\,\,\Big[ \xi_{ij} - \frac{n-2}{4( n - 1 )} \de_{ij} \Big]
\big< E^{ij} ( \Phi^{\prime}, g^{\prime} ) \big>_{J} + \,...\,\,,
\eeq
where the ellipsis denotes terms that do not contribute in the
one-loop approximation. Setting $\xi_{ij}$ to the special value
(\ref{xin}) and taking the limit $n \to 4$, we obtain
\beq
\mathcal{C}_4 \,\bar{\Ga}^{(1)}_{\text{div}} \,=\, 0\,.
\eeq
also for the model described by the classical
action~\eqref{eq:CQGwMF}. The only explicit one-loop quantum
calculation for this model was performed in Ref.~\cite{BuSh86},
where the nonconformal structures $R^2$ and
$\xi_{ij} R \phi^{i} \phi^{j}$ were eliminated through a special
conformal reparametrization procedure. The results of the present
section show that these terms are absent from the divergent part
of the one-loop effective action and, therefore, no additional field
redefinitions are required.

\section{Implications for the conformal anomaly and
induced action}
\label{sec7}

Let us briefly review the consequences of the general structure of
renormalization established in the previous sections. The
theories with conformal symmetry can be classified, in the first
place, by whether the background fields include only the metric or
also additional fields, such as matter, torsion, etc. Independent
of this separation, the possible one-loop counterterms can be
classified into three groups: \ \textit{(i)} legitimate conformal
invariants ($c$-terms). Examples include the Weyl tensor square,
electromagnetic or Yang-Mills terms, extended kinetic terms for
scalar fields, scalar self-interaction, etc.,
\beq
C^2
,\quad
F_{\mu\nu}^2
,\quad
X = \big(\pa \ph\big) + \frac16 R\ph^2
,\quad \ph^4
\,, ...\,\,.
\label{c-terms}
\eeq
Let us agree that these terms appear in the divergences in a
linear combination
\beq
Y(g, \Phi)\,=\,
\be_2 C^2
+ \be_x X
+ \be_\la \ph^4
+ \be_e F_{\mu\nu}^2\,+\,...\,\,,
\label{Y}
\eeq
where $\Phi^A$ are fields with conformal weight $d_A$, as in
Eqs.~(\ref{inftrans}) and (\ref{weights}), and the coefficients
are the corresponding beta functions. For the $c$-terms, the
transformation rule is universal,
\beq
&&
\int \rd^4x \sqrt{-g}\,\, Y(g ,\, \Phi )
\,=\,
\int \rd^4x \sqrt{- \bar{g}} \,\,Y( \bar{g} ,\, \bar{\Phi} )\,,
\label{Ytrans}
\eeq
with the conformal transformations defined in (\ref{conftrans}).

The remaining two groups form
$N$-terms, as they satisfy the conformal Noether identity but
are not conformally invariant. These groups are composed of
\textit{(ii)} the Gauss-Bonnet term; and \textit{(iii)} total
derivative terms $\na_\mu \th_{\,k}^{\mu}$, including the
metric-dependent $\Box R$. In the case of a purely metric background,
this term is unique. However, in the presence of other fields, the
list of total derivative terms may be much longer
\cite{Asorey2022,AtA}.

The extended version of the Gauss-Bonnet term (\ref{E4}) has
a simple transformation law
\beq
\sqrt{-g}\,\biggl(E_4-\frac23\,{\Box} R\biggr)
\,=\,
\sqrt{-g}\,\biggl({\bar E_4}-\frac23\,{\bar \Box} {\bar R}
+ 4{\bar \De_4}\si \biggr)\,,
\label{119}
\eeq
where $\De_4$ is the fourth-order Hermitian conformally invariant
operator \cite{FrTs-superconf,Paneitz} acting on a conformally
invariant scalar field
\beq
\De_4 = \Box^2 + 2\,R^{\mu\nu}\na_\mu\na_\nu - \frac23\,R{\Box}
+ \frac13\,(\na^\mu R)\na_\mu.
\label{120}
\eeq

The divergences can be mapped to the trace anomaly
\cite{duff77}, which can be written in the general form
\beq
\big< T^\mu_{\,\mu}\big>
\,=\,
Y \,+\,\be_2 \biggl( E_4 - \frac23\,\Box R \biggr)
\,+\,\sum_k \ga'_k \na_\mu \th_k^{\,\mu}\,,
\label{T}
\eeq
however the coefficients $\ga'_k$ may differ from
the corresponding divergences, as discussed in
\cite{birdav,duff94}, and in more detail in \cite{anomaly-2004}.
The main purpose of this section is to discuss the implications
of this difference for the status of quantum conformal gravity.

Before discussing the subtle issue of total derivative terms in the
anomaly (\ref{T}), let us write down the nonlocal part of the
anomaly-induced effective action. This functional is the solution
to the equation
\beq
-\, \frac{2}{\sqrt{-g}}\,g_{\mu\nu}\,
\frac{\de\, \Ga_{\text{ind}}}{\de g_{\mu\nu}}
\,=\, \big< T^\mu_{\,\mu}\big>\,.
\label{mainequation}
\eeq
The non-local part comes from the first two terms in (\ref{T}),
and can be immediately found using Eq.~(\ref{119}), in terms of
the Green function of the operator (\ref{120}),
\beq
\sqrt{-g(x)}\,\De^x_4\,G(x,y)\,=\,\de(x,y).
\eeq
Using the described elements, and using the notation
$\int_x = \int d^4x\sqrt{-g}$, the solution for the anomaly-induced
effective action is  \cite{rie,frts84} (see also the textbook \cite{OUP}
for more details)
\beq
&&
\Ga_{\rm ind-nonloc}
\,\,=\,\,
S_c[g_{\mu\nu},\ph]
\,-\,
\frac{\be_2 }{8}\iint\limits_{x\,y}\;\biggl(E_4-\frac23\,{\Box} R\biggr)_{\hspace{-1mm}x}
G(x,y)\biggl(E_4-\frac23\,{\Box} R\biggr)_{\hspace{-1mm}y}
\nn
\\
&&
\qquad \quad\qquad \quad
+\,\,\,
\frac{1}{4} \iint\limits_{x\,y}\; Y(x)\, G(x,y)
\biggl(E_4-\frac23\,{\Box} R\biggr)_{\hspace{-1mm}y}\,,
\label{nonlocal}
\eeq
where $S_{c}$ is an integration constant of Eq.~(\ref{mainequation}),
i.e., an unknown conformal functional.
Expression (\ref{nonlocal}) can be rewritten using two auxiliary scalar
fields $\chi$ and $\psi$ \cite{a}, as
\beq
&&
\Gamma_{\rm  ind-aux}
\,=\,
S_c[g_{\mu\nu},\ph]
\,+\, \frac12  \int_{x}\biggl\{\chi\Delta_4\chi - \psi\Delta_4\psi
\nn
\\
&&
\qquad\qquad\qquad
+\,\,
\sqrt{\be_2}\,\chi \bigg(E_4-\frac23\Box R - \frac{1}{\be_2}Y\bigg)
\,+\, \frac{1}{\sqrt{\be_2}}\,\psi Y\bigg\}.
\label{aux_fields}
\eeq
It is remarkable that the nonlocal part of the effective action in $4D$
can be obtained in such a universal form, even independently of the
presence of arbitrary background fields.

The situation is more complicated for the total derivative
terms of group \textit{(iii)}. Until recently, despite the absence of a
formal proof of this, the common belief was that there exist local
covariant Lagrangians that can produce all such terms in any theory.
This means that, for each of the terms
$\na_\mu \th_k^{\,\mu}$, there exists a local action such that
\beq
-\, \frac{2}{\sqrt{-g}}\,g_{\mu\nu}
\frac{\de }{\de g_{\mu\nu}} \,\int \rd^4x \sqrt{-g}\,\mathcal{L}_k
\,=\, \na_\mu \th_k^{\,\mu}\,.
\label{thetaL}
\eeq
The confirmation of this statement includes the six-dimensional
theory, where the set of possible total derivatives is much more
extensive compared to $4D$ \cite{Bastianelli:2000}. On the other
hand, in the recent work \cite{VecScal4D} one can find an example
of a metric-scalar theory where a local solution of
Eq.~(\ref{thetaL}) does not exist. This example concerns a
modified theory of a vector field that preserves conformal
symmetry outside $4D$, using the action \cite{ETG76}.
\beq
S [ A, \psi, g ]
\,=\,-\,\dfrac{1}{4}\int \rd^Dx\sqrt{-g}\,\psi\,F^{2}_{\mu\nu}\,.
\label{act}
\eeq
It is worth noting that the conformal invariance of the
divergences holds in the theory (\ref{act}), in full
accordance with the statement of \cite{tmf} and with
Sec.~\ref{sec3} of the present work.

The consequence for quantum conformal gravity is as follows. The
analogous gravitational models of (\ref{act}) and
(\ref{act}) can be easily constructed \cite{anomaly-2004}. In the
case with an auxiliary scalar, the classical action has the form
\beq
S [ \psi, g ]
\,=\,-\,\frac{1}{2\la}\int \rd^nx \sqrt{-g}\,\psi\,C^2(n)\,,
\label{actWeyl}
\eeq
where $C^2(n)$ is the square of the Weyl tensor  (\ref{W2})
in $n$ dimensions. There are no explicit calculations for this
model, but using the main result of the previous section and the
experience gained with the vector theory, we can expect that

1.  In both cases of the theory (\ref{actWeyl}) and the
simpler $n$-dimensional extension with $\psi \to 1$, the one-loop
divergences are conformally invariant without the reparametrization
of \cite{ETG76,frts82}. In the simpler $\psi = 1$ case, this is
confirmed by direct calculations \cite{antmot,Weyl}.

2.  In the simpler case of the theory (\ref{actionCQG}) without
an auxiliary scalar, there is only one total derivative term
$\Box R$, and it can be trivially integrated by means of relation
\beq
-\, \frac{2}{\sqrt{-g}}\,g_{\mu\nu}
\frac{\de }{\de g_{\mu\nu}} \,\int \rd^4x \sqrt{-g}\,R^2
\,=\, 12 \Box R\,.
\label{R2boxR}
\eeq

3. In the case of the theory (\ref{actWeyl}), one can expect $C^2$
plus additional $c$-terms constructed from the scalar $\psi$. There
will also be a single Gauss-Bonnet term and some extra
$\psi$-dependent total derivative terms in the anomaly. At least
some of these terms will not be integrable, in analogy with the
similar vector model \cite{VecScal4D}.

Finally, let us discuss an important issue of the ambiguity in the
$\Box R$ term in the anomaly. In semiclassical conformal theories,
this term is well known to be ambiguous, and the standard
understanding of this fact is that the ambiguity is directly related
to the possibility of adding a finite $R^2$ term (sometimes called
a “finite counterterm”) to the \textit{vacuum} classical action. The
detailed analysis of regularization ambiguities in
Ref.~\cite{anomaly-2004} shows that, in both dimensional and
covariant Pauli–Villars regularizations, these ambiguities are
completely equivalent to adding a finite $R^2$ term. For example, in
the framework of dimensional regularization one can
use $d = n + \ga(4 - n)$ in the expression (\ref{W2}) for the
counterterms, and the $\Box R$ term in the anomaly acquires a
coefficient linearly dependent on the arbitrary parameter $\ga$. The
point is that adding the nonconformal $R^2$ term to the \textit{vacuum}
action does not produce any effect on the quantum fields.

The situation in conformal quantum gravity is exactly the opposite.
Adding the nonconformal $R^2$ term introduces new degrees of freedom
into the quantum theory, which then becomes nonconformal. Thus, in this
case, this is a kind of ``illegal'' operation. So, how can we deal
with the ambiguity concerning the $\Box R$ term in the anomaly?

One of the possible answers is that the symmetry under local
conformal transformations should not be regarded as exact. The
consistent approach
implies breaking this symmetry already at the classical level, while
introducing a hierarchy in the magnitudes of the coefficients of the
$C^2$ and $R^2$ terms. Then, the analysis of the renormalization
group \cite{frts82,avbar86}
shows that this hierarchy holds at the quantum level. From
this perspective, the proof of conformal renormalizability in
quantum gravity which we have given here confirms that this kind
of effective treatment of local conformal symmetry should be
consistent beyond the tree level.

\section{Conclusions}
\label{Conc}

The main result of the present work is the proof of the one-loop
renormalizability of conformal quantum gravity, for both pure gravity
and coupled to conformal matter. Our considerations are a natural
extension of the previous work \cite{tmf}  by one of the
authors, where the same statement was proved for semiclassical
gravity, when matter fields are quantized on a classical metric
background. The statement about conformal renormalizability is
nontrivial because the quantization of the gauge theory by means
of the Faddeev-Popov approach violates conformal symmetry
of the total action (\ref{acttotal}).
The conformal invariance is achieved owing to the cancellation
of the contributions of gauge-fixing and ghost terms to the Ward
identity for the conformal symmetry. This cancellation occurs
because of the BRST invariance which remains after the
Faddeev-Popov procedure. 
The case of quantum gravity required some modifications
and, as by-product, we presented a more detailed discussion 
of the analysis proposed in \cite{tmf}.

The gauge fixing and ghosts in the conformal quantum gravity are
more complicated and the analysis of conformal renormalizability
requires more efforts. The case of quantum gravity required
modifications in the scheme, but finally we arrive at the proof
for both pure quantum gravity and the theory coupled to conformal
matter.

The consistent general proof of one-loop
renormalizability provides a solid basis for the classification of
possible terms in the trace anomaly, enables one to test the results
of calculations in new, more complicated conformal theories,
and also has independent value to provide an explanation of the 
results of existing calculations. One of the advantages is that now 
we can partially predict the general results of the new theories, 
where the practical calculations were not performed yet. The result 
is also relevant for the control of the existing and future 
calculations in conformal models.

\section*{Appendix.
\ On the conformal reparametrization of the metric}

Let us briefly review the conformal reparametrization of the metric
\cite{ETG76} and its use for restoring conformal symmetry at the
quantum level \cite{fradvilk78,frts82}.
The main formula we need is the well-known transformation of the
Ricci scalar under (\ref{conftrans}), which has the form (all
formulas below are given in $4D$, although they can be generalized
to other dimensions)
\beq
R \,=\,e^{-2\si}\big[\bar{R} - 6 \bar{\Box}\si
- 6 (\bar{\na} \si)^2 \big],
\label{confR}
\eeq
where
$(\bar{\na} \si)^2 = \bar{g}^{\mu\nu} \pa_\mu \si \pa_\nu \si $.
Our first step is to construct a metric-dependent free conformal
dimensionless scalar field $P(g)$. Such a field should satisfy the
equation
\beq
\biggl(\Box - \frac16\,R\biggr) P \,=\,0.
\label{eqP}
\eeq
As a boundary condition, we require that $P=1$ in Minkowski
spacetime, where (in particular) $R=0$. Then, it is straightforward
to verify by direct substitution that the solution can be constructed
using the Green function of the conformal operator in
Eq.~(\ref{eqP}),
\beq
P(g) \,=\, 1 \,+\,\frac16 \biggl(\Box - \frac16\,R\biggr)^{-1}R.
\label{solP}
\eeq

Since $P(g)$ satisfies the scalar equation (\ref{eqP}), its
conformal transformation is (\ref{conftrans})
\beq
P \,=\, \bar{P}\,e^{-\si}\,.
\label{confP}
\eeq
Consequently, the product
\beq
\tilde{g}_{\mu\nu} \,=\, g_{\mu\nu}\,P^2(g)
\label{tilg}
\eeq
does not transform at all under (\ref{conftrans}). This means that
we have constructed a conformally invariant version of the metric
tensor. It is clear that all geometric quantities constructed from
$\tilde{g}_{\mu\nu}$ are also conformally invariant. This
applies, in particular, to the scalar curvature $\tilde{R}$. However,
this quantity is not only conformally inert. According to
(\ref{confR}) and (\ref{tilg}), the new scalar curvature is
\beq
\tilde{R} \,=\,e^{-2\rho}\big[R - 6 \Box \rho
- 6 (\na \rho)^2 \big],
\label{tildR}
\eeq
where $\rho = \log P$. Direct inspection shows that, owing
to Eq.~(\ref{eqP}), after substituting this $\rho$ we obtain
$\tilde{R} \equiv 0$, independent of the original
metric $g_{\mu\nu}$.

The conformally invariant metric $\tilde{g}_{\mu\nu}$ can be used
in different ways. For instance, in quantum theory, it is possible 
to exploit the property $\tilde{R} = 0$. The idea is to replace the 
background metric $g_{\mu\nu}$ by $\tilde{g}_{\mu\nu}$ after 
performing the calculation of, e.g., one-loop divergences. In 
quantum conformal gravity, this amounts to eliminating the 
non-invariant $R^2$ counterterm if this structure emerges in loop 
calculations. This procedure is sometimes called ``conformal 
regularization''.

The replacement $g_{\mu\nu} \to \tilde{g}_{\mu\nu}$ can be
generalized to the metric-scalar background by adding the second
replacement $\ph \to \tilde{\ph} = \ph P^{-1}$ for each scalar field
$\ph$. This procedure enables one to eliminate the nonconformal
$R\ph^2$ term \cite{BuSh86}. Indeed, the same trick can be used 
to restore conformal invariance in many cases where this invariance 
is violated.

However, in both cases of pure quantum gravity and quantum gravity
coupled to matter -- the procedure described above is not a
regularization in the usual sense, as it does not serve to isolate
divergences. Nor can it be regarded as a legitimate procedure in 
the quantum field theory (QFT) framework. The
background field method is based on treating the background fields
without restriction until the effective action is calculated. After
this, the background fields satisfy the effective equations of
motion. In the conventional QFT framework, there is no room
for ad hoc restrictions imposed on a background field.

Last, but not least, according to the main part of the present work,
there is no need to make these replacements. If the one-loop 
calculation is a conformal theory is correct, the one-loop 
divergences should be conformal without this procedure.

\section*{Acknowledgements}

\noindent
I.L.B. is thankful to A. O. Barvinsky for useful discussion about
conformal quantum gravity. T.M.S. is grateful to Funda\c{c}\~{a}o
de Amparo \`{a} Pesquisa do Estado de Minas Gerais (FAPEMIG) for
supporting his M.Sc. project, during which part of this work was
developed, and to Coordena\c{c}\~{a}o de Aperfei\c{c}oamento de
Pessoal de N\'{i}vel Superior (CAPES) for supporting his PhD project.
I.Sh. is grateful to CNPq (Conselho Nacional de Desenvolvimento
Cient\'{i}fico e Tecnol\'{o}gico, Brazil)  for the partial support
under the grant 305122/2023-1.



\end{document}